\documentclass[11pt,a4paper]{article}

\usepackage[utf8]{inputenc}
\usepackage[T1]{fontenc}
\usepackage{amsmath,amssymb}
\usepackage{graphicx}
\usepackage{booktabs}
\usepackage{hyperref}
\usepackage{xcolor}
\usepackage{algorithm}
\usepackage{algpseudocode}
\usepackage{multirow}
\usepackage{tikz}
\usetikzlibrary{arrows.meta, positioning, shapes.geometric}
\usepackage{geometry}
\usepackage[numbers]{natbib}
\usepackage{enumitem}

\hypersetup{colorlinks=true, linkcolor=blue, citecolor=blue, urlcolor=blue}

\title{Code-QA-Bench: Separating Code Reasoning from Documentation Memorization \\ in Repository-Level QA}

\author{
  Jun Zhang \and JianYing Qu \and Hanwen Du \and Zhongkai Sun \and Yehua Yang \and Qiao Zhao \\
  Baidu Inc.
}

\date{May 27, 2026}

\begin{document}
\maketitle

\begin{abstract}
We present \textbf{Code-QA-Bench}, a fully automated framework for synthesizing repository-level code understanding benchmarks that separates genuine code comprehension from documentation recall and pretraining memorization.
The framework makes two methodological contributions: (1)~an \emph{answer-first} generation pipeline where a tool-equipped agent explores source code to produce verified gold answers before deriving questions, ensuring every task is grounded in real code structure; and (2)~a \emph{three-condition} experimental design evaluating agents under closed-book (no repository), code-only (documentation removed), and documented (full repository) conditions, with deltas directly quantifying documentation utility and memorization.

We generate 528 code-derivable and 100 doc-dependent tasks across 10 Python repositories from SWE-Bench, scored by an LLM judge on accuracy, completeness, and specificity.
Experiments on four frontier models reveal that code access is the dominant factor (+0.23 mean gain over closed-book), documentation provides modest additional benefit (+0.071 on doc-dependent tasks, $p < 0.003$), and code-only $\approx$ documented on code-derivable tasks ($\Delta < 0.01$), validating the design.
The framework is open-source and applicable to any well-documented Python repository.
\end{abstract}


\section{Introduction}
\label{sec:intro}

The dominant use of AI coding assistants in practice is not writing code from scratch, but understanding existing codebases~\citep{xia2024survey}.
Code understanding is broad: it includes reasoning about the input/output behavior of a function, locating the code responsible for a given feature, tracing control flow across modules, and building enough context to confidently propose a bug fix.
Yet the most prominent benchmarks for AI coding agents (HumanEval~\citep{chen2021codex}, MBPP~\citep{austin2021program}, and SWE-Bench~\citep{jimenez2024swebench}) focus on code generation or issue resolution, leaving code \emph{comprehension} under-evaluated.

This gap matters.
An agent that can generate correct patches may still fail to explain why a function behaves a certain way, or to locate the right file for a given concept in a 500-file repository.
Code understanding is a prerequisite for safe and reliable code modification, and measuring it independently provides signal that generation benchmarks miss.
CRUXEval~\citep{gu2024cruxeval} highlighted this by showing that code generation ability does not imply code reasoning ability, but operates only at the function level.
RepoReason~\citep{li2025reporeason} extended reasoning evaluation to the repository level, revealing that aggregating information across files is the primary cognitive bottleneck for frontier models.
Recent work on repository-level code QA~\citep{peng2025sweqa, cai2025sweqapro} has made progress on question taxonomies and difficulty calibration, but evaluates agents on repositories with full documentation intact, making it difficult to separate genuine code reading from documentation recall or pretraining memorization.

We introduce \textbf{Code-QA-Bench}, a benchmark designed around four principles:
(1)~a \emph{three-condition experimental design} evaluating each task under closed-book (no repository), code-only (documentation removed), and documented (full repository) conditions, with deltas directly quantifying documentation utility and memorization;
(2)~\emph{documentation removal as an environment-level control}, removing all docstrings, comments, and documentation files to force code-structural reasoning while retaining semantic signal from identifiers, imports, and type annotations (Section~\ref{sec:code-only-semantics});
(3)~\emph{answer-first task generation} where a tool-equipped agent explores source code to produce verified gold answers before deriving questions, ensuring every task is grounded in real code structure; and
(4)~\emph{unified continuous scoring} via an LLM judge on three axes (accuracy, completeness, specificity), each 0--5, normalized to a single 0--1 score.

Code-QA-Bench uses 10 Python repositories from SWE-Bench~\citep{jimenez2024swebench} (Table~\ref{tab:repos}), pinned to specific commits for reproducibility.
Tasks follow a four-type taxonomy (What/Why/Where/How) with 12 subtypes from SWE-QA~\citep{peng2025sweqa}, balanced per repository.
The entire pipeline is fully automated and \emph{repo-agnostic}: it can be applied to any well-documented Python repository without modification.

\section{Related Work}
\label{sec:related}

\paragraph{Function-level and repository-level code benchmarks.}
Early code understanding benchmarks operate at the function level: CRUXEval~\citep{gu2024cruxeval}, CodeQA~\citep{liu2021codeqa}, CS1QA~\citep{lee2022cs1qa}, LiveCodeBench~\citep{jain2024livecodebench}, and EvalPlus~\citep{liu2024evalplus}.
None capture the cross-file reasoning demands of real-world development.
At the repository level, SWE-Bench~\citep{jimenez2024swebench} and its extensions~\citep{aleithan2024swebenchplus, xie2025featurebench, chou2025autocodebench} evaluate code \emph{generation} via patches, while others address cross-file completion~\citep{ding2024crosscodeeval, liu2023repobench}, lifecycle tasks~\citep{li2024deveval, li2024devbench}, and agent architectures~\citep{yang2024sweagent, xia2024agentless}.
In all cases, code understanding is a means to generation, not the measured outcome.

\paragraph{Repository-level code QA.}
SWE-QA~\citep{peng2025sweqa} is the first large-scale repository-level code QA benchmark (576 questions, 12 repositories, 4-type taxonomy).
SWE-QA-Pro~\citep{cai2025sweqapro} improves on it by selecting 26 long-tail repositories and applying difficulty calibration to filter questions answerable without tools, demonstrating that many existing items do not require repository interaction.
Other work evaluates code reasoning through assertion verification~\citep{li2025reporeason}, execution prediction~\citep{liu2024codemind}, multi-dimensional frameworks~\citep{yan2024codescope}, and automated benchmark synthesis~\citep{jain2024r2e}.
Our work addresses the memorization problem from a different angle: rather than filtering easy questions or selecting unfamiliar repositories, we \emph{directly measure} memorization via three evaluation conditions and adopt SWE-QA's taxonomy with balanced distribution and an answer-first generation pipeline.

\paragraph{Evaluation methodology.}
We use rubric-guided LLM-as-judge evaluation~\citep{zheng2023judging, li2024crowdsource} with a 3-axis rubric emphasizing concreteness.
Our documentation-removal approach to contamination control relates to CodeSearchNet's~\citep{husain2019codesearchnet} separation of code from natural-language descriptions, and complements temporal methods~\citep{jain2024livecodebench} and difficulty calibration~\citep{cai2025sweqapro} by removing the natural-language layer most susceptible to memorization.

\section{Benchmark Design}
\label{sec:design}

\subsection{Repository Selection and Pinning}
\label{sec:repos}

We select 10 of the 12 SWE-Bench Python repositories~\citep{jimenez2024swebench} (Table~\ref{tab:repos}), excluding Flask and Requests due to insufficient documentation volume.
Unlike SWE-QA-Pro~\citep{cai2025sweqapro}, which avoids popular repositories, we retain well-known projects and address memorization via the three-condition design (Section~\ref{sec:conditions}).

Each repository is pinned to a specific git commit SHA for reproducibility.
Task counts are allocated by size tier: large ($>$1{,}500 \texttt{.py} files) receive 72 tasks, medium (200--1{,}000) receive 48, and small ($<$200) receive 24, yielding 528 total.
The pipeline makes no assumptions specific to these repositories and can be applied to any well-documented Python project.

\begin{table}[t]
\centering
\caption{Benchmark repositories, drawn from SWE-Bench~\citep{jimenez2024swebench}. Task counts are allocated by repository size tier.}
\label{tab:repos}
\begin{tabular}{llclc}
\toprule
\textbf{Repository} & \textbf{Domain} & \textbf{.py Files} & \textbf{Commit SHA} & \textbf{Tasks} \\
\midrule
django         & Web framework            & 2{,}894 & \texttt{856c9153...} & 72 \\
pylint         & Static code analysis      & 2{,}366 & \texttt{b715af6c...} & 72 \\
sympy          & Symbolic mathematics      & 1{,}589 & \texttt{693a559a...} & 72 \\
\midrule
scikit-learn   & Machine learning          & 993 & \texttt{f3182980...} & 48 \\
astropy        & Astronomy library         & 987 & \texttt{30686862...} & 48 \\
matplotlib     & Plotting / visualization  & 913 & \texttt{cc6cead6...} & 48 \\
sphinx         & Documentation generator   & 774 & \texttt{cc7c6f43...} & 48 \\
pytest         & Testing framework         & 262 & \texttt{8ecf49ec...} & 48 \\
xarray         & N-D labeled arrays        & 237 & \texttt{92601de1...} & 48 \\
\midrule
seaborn        & Statistical visualization & 151 & \texttt{32088bbc...} & 24 \\
\midrule
\multicolumn{4}{r}{\textbf{Total}} & \textbf{528} \\
\bottomrule
\end{tabular}
\end{table}

\subsection{Documentation Removal}
\label{sec:doc-removal}

To isolate code comprehension from documentation recall, we create a code-only version of each repository that removes three categories of natural-language content:

\begin{enumerate}[leftmargin=*]
  \item \textbf{Docstrings.} We use Python's Abstract Syntax Tree (AST) module to identify docstrings (string literals that are the first statement in a module, class, or function body).
  These are removed from the source.
  When removing a docstring would leave an empty body (e.g., a function whose only statement is its docstring), a \texttt{pass} statement is inserted to maintain syntactic validity.

  \item \textbf{Comments.} We use Python's \texttt{tokenize} module to identify comment tokens.
  Full-line comments (where only whitespace precedes the \texttt{\#}) are removed entirely.
  Inline comments are truncated at the \texttt{\#} position.

  \item \textbf{Documentation files.} The directories \texttt{docs/}, \texttt{doc/}, and \texttt{.git/} are deleted.
  Files matching \texttt{README*}, \texttt{*.md}, \texttt{*.rst}, \texttt{CHANGELOG*}, and \texttt{CONTRIBUTING*} are removed.
\end{enumerate}

The removal is deterministic and preserves all executable code, imports, type annotations, and string literals.
Unlike SWE-QA-Pro's question-level difficulty calibration~\citep{cai2025sweqapro}, our environment-level approach ensures \emph{all} questions require genuine code reading.

\subsection{Task Categories}
\label{sec:categories}

We adopt the four-type taxonomy from SWE-QA~\citep{peng2025sweqa}, with 12 Level-2 subtypes derived from their analysis of 77{,}100 real GitHub issue questions:

\begin{itemize}[leftmargin=*]
  \item \textbf{What} (factual inquiry): architecture exploration, concept definition, dependency tracing. Example: ``What components make up the caching subsystem?''
  \item \textbf{Why} (causal explanation): design rationale, purpose exploration, performance. Example: ``Why does the ORM use lazy evaluation for querysets?''
  \item \textbf{Where} (location identification): data/control flow, feature location, identifier location. Example: ``Where is the request routing logic implemented?''
  \item \textbf{How} (procedural explanation): system design, algorithm implementation, API/framework support. Example: ``How does the template engine resolve variable lookups?''
\end{itemize}

Each repository's tasks are split evenly across the four Level-1 categories (25\% each) via round-robin assignment during chunk selection.
Within each category, subtypes are distributed evenly via a secondary round-robin.
This balanced design ensures that no category dominates and enables per-category analysis of model strengths and weaknesses.
We hypothesize that \emph{Why} questions, which often require understanding design rationale typically found in comments and documentation, will show the largest gap between code-only and documented conditions.

Categories serve as metadata for breakdown analysis but do not affect the overall score.

\section{Task Generation Pipeline}
\label{sec:generation}

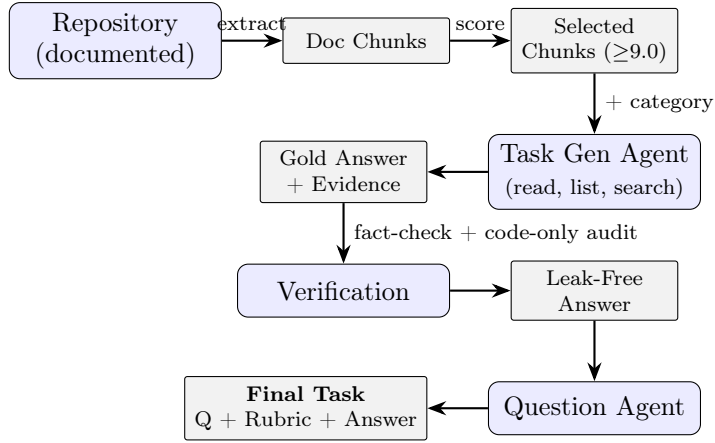
\begin{figure}[t]
\centering
\begin{tikzpicture}[
    node distance=0.6cm and 0.4cm,
    stage/.style={draw, rounded corners, fill=blue!8, minimum width=2.8cm, minimum height=0.7cm, align=center, font=\small},
    data/.style={draw, rounded corners=1pt, fill=gray!10, minimum width=2.2cm, minimum height=0.6cm, align=center, font=\scriptsize},
    arr/.style={-{Stealth[length=2.5mm]}, thick},
    label/.style={font=\scriptsize, midway, above},
]
\node[stage] (repo) {Repository\\(documented)};
\node[data, right=0.8cm of repo] (chunks) {Doc Chunks};
\node[data, right=0.8cm of chunks] (selected) {Selected\\Chunks ($\geq$9.0)};

\draw[arr] (repo) -- node[label] {extract} (chunks);
\draw[arr] (chunks) -- node[label] {score} (selected);

\node[stage, below=0.8cm of selected] (agent) {Task Gen Agent\\{\scriptsize (read, list, search)}};
\node[data, left=0.8cm of agent] (gold) {Gold Answer\\+ Evidence};

\draw[arr] (selected) -- node[font=\scriptsize, right] {+ category} (agent);
\draw[arr] (agent) -- (gold);

\node[stage, below=0.8cm of gold] (verify) {Verification};
\node[data, right=0.8cm of verify] (verified) {Leak-Free\\Answer};

\draw[arr] (gold) -- node[font=\scriptsize, right] {fact-check + code-only audit} (verify);
\draw[arr] (verify) -- (verified);

\node[stage, below=0.8cm of verified] (qgen) {Question Agent};
\node[data, left=0.8cm of qgen] (task) {\textbf{Final Task}\\Q + Rubric + Answer};

\draw[arr] (verified) -- (qgen);
\draw[arr] (qgen) -- (task);
\end{tikzpicture}
\caption{The answer-first task generation pipeline. Documentation chunks are extracted and scored (Table~\ref{tab:chunk-scoring}); chunks $\geq 9.0$ are selected. The Task Generation Agent explores source code to produce a gold answer, which is verified and code-only audited (\S\ref{sec:strip-verify}); a question and rubric are then derived from the verified answer.}
\label{fig:pipeline}
\end{figure}

The task generation pipeline follows an \emph{answer-first} design: we generate code-grounded gold answers before deriving questions.
This inverts the typical approach used in SWE-QA~\citep{peng2025sweqa} (question templates from taxonomy) and SWE-QA-Pro~\citep{cai2025sweqapro} (question synthesis from issue clusters), and ensures every question has a verifiable, specific reference answer rooted in actual source code.
The pipeline consists of five stages (Figure~\ref{fig:pipeline}).

\subsection{Stage 1: Documentation Chunk Extraction}
\label{sec:chunks}

We extract documentation from the documented (full) repository from three sources:

\begin{itemize}[leftmargin=*]
  \item \textbf{README chunks.} README files are split by \texttt{\#\#} markdown headings. Each heading--body pair becomes a chunk.
  \item \textbf{Documentation file chunks.} Files under \texttt{docs/} and \texttt{doc/} (Markdown and reStructuredText) are split by headings. For RST files, we detect underline-style headings (\texttt{====}, \texttt{----}, \texttt{\~{}\~{}\~{}}, etc.) in addition to Markdown-style headings, using whichever method finds more sections. This is critical for RST-heavy projects like Django (724 doc files, mostly RST).
  \item \textbf{Docstring chunks.} All Python files are AST-parsed to extract docstrings. Rather than aggregating all docstrings from a single module into one mega-chunk, we emit one chunk per top-level class (with its methods' docstrings grouped) and one chunk per top-level function. Module-level docstrings become their own chunks. This finer granularity produces more focused, class-specific chunks that ground better task generation.
\end{itemize}

RST directive code references (\texttt{:func:}, \texttt{:class:}, \texttt{:meth:}, \texttt{:mod:}) are recognized alongside backtick identifiers when scoring chunks (Section~\ref{sec:scoring-chunks}), ensuring that RST-formatted documentation is not disadvantaged relative to Markdown.

\subsection{Stage 2: Chunk Scoring and Selection}
\label{sec:scoring-chunks}

Each chunk receives a deterministic heuristic score based on multiple signals organized into three groups: content quality, path-based, and content bonuses/penalties.

\begin{table}[h]
\centering
\caption{Chunk scoring heuristic. Signals are grouped into baseline content signals, path-based penalties that demote non-core code, and content-quality bonuses/penalties that reward structured documentation and penalize low-information chunks.}
\label{tab:chunk-scoring}
\small
\begin{tabular}{p{0.52\textwidth}rp{0.30\textwidth}}
\toprule
\textbf{Signal} & \textbf{Pts} & \textbf{Rationale} \\
\midrule
\multicolumn{3}{l}{\emph{Baseline content signals}} \\
Length 200--8000 chars        & $+2$ & Enough detail, not too noisy \\
Code references (backticks, dotted paths) & $+3$ & Grounds in specific code \\
Behavioral keywords (\emph{returns, raises, calls, when, if}) & $+2$ & Describes runtime behavior \\
Source is docstring           & $+1$ & More code-proximal than README \\
Generic heading (\emph{install, license, contributing}) & $-2$ & Not code-understanding \\
\midrule
\multicolumn{3}{l}{\emph{Path-based penalties}} \\
Test file (\texttt{/test} or \texttt{test\_} in path) & $-2$ & Test behavior, not design \\
Vendored code (\texttt{extern}, \texttt{vendor}) & $-2$ & Third-party, not repo's own \\
Example/gallery in path & $-1$ & Tutorial-like \\
\midrule
\multicolumn{3}{l}{\emph{Content-quality bonuses and penalties}} \\
Structured sections (\emph{Params} + \emph{Returns}) & $+1$ & Strong QA grounding \\
Doctest examples (\texttt{>\kern-.1em>\kern-.1em>}) & $+1$ & Concrete usage \\
Deep module path ($\geq$3 levels) & $+1$ & Implementation-focused \\
Stacked one-liners ($>$8, avg $<$80 chars) & $-2$ & Low information density \\
RST directive-heavy ($>$30\% of lines) & $-1$ & Mostly markup \\
\bottomrule
\end{tabular}
\end{table}

Chunks are ranked by score and the top $N$ per repository are selected (where $N$ follows the size tier in Table~\ref{tab:repos}), with at most 2 chunks per source file.
Categories are assigned via round-robin to ensure 25\% per category within each repository.

\subsection{Stage 3: Agentic Gold Answer Generation}
\label{sec:answer-gen}

For each selected chunk, the \emph{Task Generation Agent} produces a gold answer grounded in actual source code.
The agent operates in a multi-turn tool-use loop with three read-only tools: \texttt{read\_file}, \texttt{list\_directory}, and \texttt{search\_code} (regex over all \texttt{.py} files).
A 200K-character context budget prevents overflow.

The agent receives the documentation chunk as a \emph{topic guide} but must go beyond it: the system prompt requires tracing at least one level deeper (callees, parent classes, or related modules) and including facts \emph{not} in the documentation.
Per-category guidance steers exploration strategy (e.g., \emph{Where} traces call chains; \emph{How} follows implementations line by line).
The agent outputs structured JSON with a detailed answer, key file paths, and $\geq$3 code evidence items citing specific files and functions.
Key files are validated against the filesystem; missing paths trigger a re-prompt with correction hints.

\subsection{Stage 4: Verification and Code-Only Verification}
\label{sec:verification}
\label{sec:strip-verify}

Two automated quality gates filter the generated answers.
First, a \textbf{fact-check} LLM call verifies the answer against the documentation chunk; tasks with significant contradictions are dropped (Pass/Warn/Fail).
Second, a \textbf{code-only verification} pass addresses \emph{doc-leakage}: since agents are evaluated on repositories with documentation removed, the gold answer must not contain claims only recoverable from documentation.
An LLM auditor receives the gold answer alongside the code-only key files and classifies each claim as Keep (verifiable from code), Remove (documentation-dependent), or Rewrite (partially verifiable).
The revised answer replaces the original, and the doc-leakage fraction is recorded per task (full prompt in Appendix~\ref{app:strip-verify-prompt}).

\subsection{Dual Task Set Design}
\label{sec:dual-tasks}

Code-only-verified tasks should yield similar scores under code-only and documented conditions; we call these \textbf{code-derivable tasks} (528 tasks).
To measure documentation utility, we additionally generate 100 \textbf{doc-dependent tasks} (10 per repository) using a different branch: gold answers are produced in a single LLM turn from documentation only (no code tools), are \emph{not} code-only-verified, and intentionally require documentation to fully address.
Code-derivable tasks validate the design (code-only $\approx$ documented expected); doc-dependent tasks quantify documentation utility (documented $>$ code-only expected).

\subsection{Stage 5: Question Generation}
\label{sec:question-gen}

In its second phase, the Task Generation Agent derives a natural question from the verified gold answer.
The agent uses the same tool set (allowing it to re-check details against the code) but operates under a different system prompt and output schema: it produces a short ($\sim$25 words), conversational question and a rubric of key points.
The agent is instructed to write the kind of question a developer would actually ask, avoiding academic-sounding phrasing like ``trace through'' or ``provide a comprehensive analysis.''

The two-phase design (answer first, question second) ensures that answer quality is established and verified before a question is derived from it, reducing the risk of ill-posed or unanswerable questions.
The rubric produced in this phase is used by the LLM judge during evaluation.

\section{Evaluation}
\label{sec:evaluation}

\subsection{Three-Condition Experimental Design}
\label{sec:conditions}

All 528 tasks are evaluated under three conditions for every model:

\begin{table}[h]
\centering
\caption{Three experimental conditions. All tasks are evaluated under all conditions.}
\label{tab:conditions}
\begin{tabular}{lll}
\toprule
\textbf{Condition} & \textbf{Agent Input} & \textbf{Measures} \\
\midrule
Closed-book & Question only, no repo access & Memorization / prior knowledge \\
Code-only (primary) & Question + code-only repo & Code-structural reasoning \\
Documented & Question + full repo with docs & Code reasoning + documentation \\
\bottomrule
\end{tabular}
\end{table}

The key analysis deltas are:
\begin{itemize}[leftmargin=*]
  \item \textbf{Documented $-$ Code-only} = documentation utility for AI code comprehension.
  \item \textbf{Code-only $-$ Closed-book} = genuine contribution of code reading beyond memorization.
  \item \textbf{Per-category breakdown}: we hypothesize that \emph{Why} questions benefit most from documentation (rationale in comments), while \emph{Where} questions may not (requires code tracing regardless).
\end{itemize}

Tasks with high closed-book scores are flagged as potentially contaminated.
The code-only condition is the primary benchmark metric; the other two conditions provide diagnostic context.

\subsection{Scoring}
\label{sec:scoring}

Every agent answer is evaluated by an LLM judge on three axes:

\begin{itemize}[leftmargin=*]
  \item \textbf{Accuracy} (0--5): Are the factual claims in the answer correct? Points are deducted for incorrect statements about the code.
  \item \textbf{Completeness} (0--5): Does the answer cover all key points in the rubric? $5 =$ all points covered, $0 =$ none.
  \item \textbf{Specificity} (0--5): Does the answer reference specific files, functions, classes, or code patterns? $5 =$ very specific with file/function names, $0 =$ entirely vague.
\end{itemize}

The per-question score is the mean of the three axes, normalized to $[0, 1]$:

\begin{equation}
  s_i = \frac{\text{accuracy}_i + \text{completeness}_i + \text{specificity}_i}{15}
\end{equation}

The overall benchmark score is the simple mean across all $N$ questions:

\begin{equation}
  S = \frac{1}{N} \sum_{i=1}^{N} s_i
\end{equation}

Our 3-axis rubric differs from the 5-dimension evaluation used in SWE-QA~\citep{peng2025sweqa} and SWE-QA-Pro~\citep{cai2025sweqapro} (correctness, completeness, relevance, clarity, reasoning).
We deliberately omit ``clarity'' and ``reasoning'' as evaluation axes because in a code-only setting, we care more about whether the agent found the right code and reported accurate facts than about the prose quality of its explanation.

\textbf{Note on specificity saturation.}
When agents have repository access (code-only or documented conditions), specificity saturates near 5.0 ($\geq$4.78 for all models), because agents naturally reference concrete file paths and function names.
In practice, the composite score in agentic conditions is primarily driven by accuracy and completeness; specificity serves as a discriminator only in the closed-book condition where models cannot ground their answers in specific code locations.
We add ``specificity'' as a dedicated axis because referencing concrete file paths and function names is the primary evidence that an agent genuinely navigated the codebase rather than providing a plausible-sounding but vague response.

The simple mean treats every question equally.
Category breakdowns (What/Why/Where/How) and per-condition breakdowns (closed-book/code-only/documented) remain available for diagnostic analysis.

\subsection{Judge Model}
\label{sec:judge-model}

We use GPT-5.4 as the LLM judge, deliberately choosing a model from a different provider and model family than the task generation model (Claude Opus 4.6) to mitigate self-evaluation bias.
The judge receives: the original question, the gold reference answer, the rubric, and the agent's answer.
It returns a structured JSON response with the three scores and a brief explanation.

To mitigate known biases in LLM judges~\citep{zheng2023judging}, including position bias and verbosity bias, the judge prompt uses a structured rubric with explicit scoring criteria.
SWE-QA~\citep{peng2025sweqa} additionally anonymizes candidates and randomizes answer order; we adopt similar practices.

For the \emph{relative} comparisons that form our primary analysis (condition deltas, model rankings), judge consistency matters more than absolute calibration.

\section{Experimental Setup}
\label{sec:experiments}

\subsection{Agent Architecture}
\label{sec:agents}

Code-QA-Bench ships with a built-in evaluation agent that reuses the same infrastructure as the Task Generation Agent (Section~\ref{sec:answer-gen}): the same tool definitions (\texttt{read\_file}, \texttt{list\_directory}, \texttt{search\_code}), the same tool execution engine, and the same multi-turn loop (up to 60 turns).
The evaluation agent uses condition-specific system prompts:

\begin{itemize}[leftmargin=*]
  \item \textbf{Code-only condition}: The agent is told that ``the repository has been stripped of all docstrings, comments, and documentation files'' and must ``derive understanding from code structure, naming conventions, control flow, and type signatures alone.''
  \item \textbf{Documented condition}: The agent is told that ``the repository includes full docstrings, comments, and documentation files'' and is instructed to ``start by searching for relevant documentation (README, docs/, CHANGELOG, .rst/.md files)'' before diving into code.
\end{itemize}

Since gold answers were produced by an agent with exactly these tools, any performance gap reflects code understanding ability rather than tooling differences.
The benchmark also exposes a pluggable agent interface: any async callable matching \texttt{(repo\_path, question) -> answer} can substitute the built-in agent.

\subsection{Models Evaluated}

\begin{table}[h]
\centering
\caption{Models evaluated. All use the same built-in agent with identical tool access (60 max turns, 4096 max tokens per response).}
\label{tab:models}
\begin{tabular}{llll}
\toprule
\textbf{Model} & \textbf{API Identifier} & \textbf{Provider} & \textbf{Context} \\
\midrule
Claude Opus 4.6 & \texttt{claude-opus-4-6} & Anthropic & 200K \\
DeepSeek-V4-Pro & \texttt{deepseek-v4-0324} & DeepSeek & 128K \\
Kimi-K2.6 & \texttt{kimi-k2.6-0528} & Moonshot AI & 128K \\
Gemini-3.1-Pro & \texttt{gemini-3.1-pro-preview} & Google & 1M \\
\bottomrule
\end{tabular}
\end{table}

Claude Opus 4.6 also serves as the task generation model, creating a potential circularity.
We mitigate this via three design choices: (1)~a separate GPT-5.4 judge (different provider, no self-evaluation); (2)~generation uses documented code while evaluation uses code-only code; (3)~rubric-based scoring constrains evaluation to factual key points.
Empirically, the bias is negligible on the primary metric: DeepSeek (0.892) matches Claude (0.891) on code-derivable tasks in the code-only condition despite having no role in task generation.
All models are evaluated through a unified inference gateway with identical tool definitions and system prompts.

\subsection{Evaluation Protocol}

Each model is evaluated under all three conditions (Section~\ref{sec:conditions}).
Generation parameters: \texttt{max\_tokens}${}=4096$, \texttt{max\_turns}${}=60$, \texttt{context\_budget}${}=200{,}000$ characters.
All agent and judge calls include retry logic with exponential backoff.

\section{Results}
\label{sec:results}

We present results on both the code-derivable (528 tasks) and doc-dependent (100 tasks) task sets across all four models and three conditions.

All reported $p$-values use paired bootstrap (10,000 resamples) with Bonferroni correction ($\alpha_{\text{adj}}=0.003$ for 16 comparisons); full statistical details and 95\% confidence intervals are in Appendix~\ref{app:bootstrap}.
Score distributions in the code-only condition are left-skewed with 5--12\% of tasks at ceiling ($\geq$0.95) and 1--15\% below 0.70; closed-book distributions have 87--97\% of tasks below 0.70, confirming that most tasks genuinely require code access.

\subsection{Doc-Dependent Task Results}
\label{sec:results-doc-dep}

Table~\ref{tab:doc-dep-results} shows results on the 100 doc-dependent tasks.
This task set is designed to measure documentation utility: gold answers contain information only available in documentation, so agents with documentation access (documented condition) should outperform those without.

\begin{table}[h]
\centering
\caption{Results on doc-dependent tasks (100 tasks). Score is the normalized mean of accuracy, completeness, and specificity (each 0--5, normalized to 0--1). $\Delta_{\text{doc}}$ = documented $-$ code-only measures documentation utility. All $\Delta_{\text{doc}}$ values are statistically significant ($p < 0.003$, paired bootstrap test, 10K resamples; full CIs in Appendix~\ref{app:bootstrap}).}
\label{tab:doc-dep-results}
\begin{tabular}{lcccc}
\toprule
\textbf{Model} & \textbf{Closed-book} & \textbf{Code-only} & \textbf{Documented} & $\boldsymbol{\Delta_{\text{doc}}}$ \\
\midrule
Claude Opus 4.6 & 0.682 & 0.873 & \textbf{0.953} & +0.080 \\
Kimi-K2.6 & 0.639 & 0.859 & 0.950 & +0.091 \\
DeepSeek-V4-Pro & 0.557 & 0.867 & 0.928 & +0.061 \\
Gemini-3.1-Pro & 0.636 & 0.840 & 0.893 & +0.053 \\
\midrule
\textbf{Mean} & 0.629 & 0.860 & 0.931 & +0.071 \\
\bottomrule
\end{tabular}
\end{table}

Three clear patterns emerge:

\paragraph{Documentation provides consistent, measurable benefit.}
Across all four models, the documented condition outperforms code-only by +0.053 to +0.091 (mean +0.071).
This gap is consistent across model families and confirms that agents can effectively use documentation when questions genuinely require it.
The gap is moderate rather than dramatic, suggesting that much of the answer can be partially inferred from code structure, but documentation provides the final details (design rationale, deprecation warnings, edge case caveats) that push scores from $\sim$0.86 to $\sim$0.93.

\paragraph{Code access dominates over memorization.}
The code-only$-$closed-book gap (mean +0.231) is three times larger than the documented$-$code-only gap (mean +0.071), demonstrating that \emph{reading code} contributes far more to understanding than documentation alone.
Even for questions explicitly designed to require documentation, code structure provides substantial signal.

\paragraph{Significant parametric knowledge in closed-book.}
Models score 0.56--0.68 without any repository access, indicating that frontier models have memorized substantial information about these well-known Python libraries during pretraining.
Claude Opus achieves the highest closed-book score (0.682), consistent with its larger pretraining corpus.

\subsection{Code-Derivable Task Results}
\label{sec:results-code-derivable}

Table~\ref{tab:code-derivable-results} shows results on the 528 code-derivable tasks.
These tasks are code-only-verified: gold answers are recoverable from code structure alone.

\begin{table}[h]
\centering
\caption{Results on code-derivable tasks (528 tasks). Tasks are code-only-verified, so code-only $\approx$ documented is expected. $\Delta_{\text{doc}}$ is near zero for three models ($p > 0.03$); Gemini's negative $\Delta$ is marginally significant ($p = 0.004$, just above Bonferroni-adjusted $\alpha=0.003$); Claude's positive $\Delta$ reflects generator advantage ($p < 0.001$). Full bootstrap CIs in Appendix~\ref{app:bootstrap}.}
\label{tab:code-derivable-results}
\begin{tabular}{lcccc}
\toprule
\textbf{Model} & \textbf{Closed-book} & \textbf{Code-only} & \textbf{Documented} & $\boldsymbol{\Delta_{\text{doc}}}$ \\
\midrule
Claude Opus 4.6 & 0.560 & 0.891 & \textbf{0.918} & +0.026 \\
DeepSeek-V4-Pro & 0.442 & 0.892 & 0.899 & +0.007 \\
Kimi-K2.6 & 0.514 & 0.873 & 0.882 & +0.008 \\
Gemini-3.1-Pro & 0.482 & 0.772 & 0.755 & $-$0.018 \\
\midrule
\textbf{Mean} & 0.500 & 0.857 & 0.864 & +0.006 \\
\bottomrule
\end{tabular}
\end{table}

The key validation: \textbf{code-only $\approx$ documented on code-derivable tasks} (mean $\Delta_{\text{doc}}$ = +0.007).
For DeepSeek and Kimi, the documented$-$code-only delta is not statistically significant ($p > 0.03$, paired bootstrap), confirming that code-only verification produces tasks whose answers are recoverable from code alone.
The near-zero documentation gap validates our experimental design: any gap observed on doc-dependent tasks (mean $\Delta_{\text{doc}}$ = +0.071, $p < 0.003$ for all models) is genuinely attributable to documentation utility, not an artifact of the evaluation setup.
All code-only$-$closed-book deltas are highly significant ($p < 0.001$), confirming that code access provides substantial information beyond parametric knowledge.

Several model-specific patterns emerge:

\paragraph{Claude Opus 4.6 leads across conditions.}
As the task generation model, Claude Opus achieves the highest scores in both code-only (0.891) and documented (0.918). Its closed-book score (0.560) is also highest, reflecting both its large pretraining corpus and the advantage of having generated the gold answers from the same codebase knowledge.

\paragraph{DeepSeek-V4-Pro: strong code reader, weak memorizer.}
DeepSeek achieves near-top code-only performance (0.892) with the lowest closed-book score (0.442), the largest gap of any model (+0.450). This suggests DeepSeek relies heavily on active code exploration rather than parametric recall.

\paragraph{Gemini shows a negative documentation effect.}
Gemini-3.1-Pro is the only model where documented (0.755) underperforms code-only (0.772), yielding $\Delta_{\text{doc}} = -0.018$ ($p = 0.004$, marginally significant at Bonferroni-adjusted $\alpha = 0.003$).
Documentation presence likely diverts the agent's exploration strategy toward README/docs files that provide no signal for code-derivable questions, consuming turns from a limited budget.

\paragraph{Closed-book scores are lower than on doc-dependent tasks.}
Mean closed-book on code-derivable (0.500) is below doc-dependent (0.629), suggesting that implementation-level details (function signatures, control flow, class hierarchies) are harder to recall from pretraining than documentation-level knowledge (design rationale, usage patterns).

\subsection{Axis-Level Analysis}
\label{sec:axis-analysis}

Table~\ref{tab:axis-analysis} breaks down the doc-dependent scores by evaluation axis.

\begin{table}[h]
\centering
\caption{Per-axis scores on doc-dependent tasks (0--5 scale). Specificity saturates near ceiling with code access; completeness shows the most variation across conditions.}
\label{tab:axis-analysis}
\begin{tabular}{llccc}
\toprule
\textbf{Model} & \textbf{Condition} & \textbf{Accuracy} & \textbf{Completeness} & \textbf{Specificity} \\
\midrule
\multirow{3}{*}{Claude Opus 4.6}
  & Closed-book & 3.05 & 3.13 & 4.05 \\
  & Code-only & 4.15 & 3.98 & 4.97 \\
  & Documented & 4.62 & 4.68 & 5.00 \\
\midrule
\multirow{3}{*}{DeepSeek-V4-Pro}
  & Closed-book & 2.43 & 2.38 & 3.54 \\
  & Code-only & 4.22 & 3.95 & 4.97 \\
  & Documented & 4.53 & 4.55 & 4.99 \\
\midrule
\multirow{3}{*}{Kimi-K2.6}
  & Closed-book & 3.02 & 2.75 & 3.82 \\
  & Code-only & 4.33 & 3.71 & 4.85 \\
  & Documented & 4.80 & 4.46 & 4.99 \\
\midrule
\multirow{3}{*}{Gemini-3.1-Pro}
  & Closed-book & 3.28 & 2.71 & 3.55 \\
  & Code-only & 4.13 & 3.65 & 4.82 \\
  & Documented & 4.58 & 4.18 & 4.78 \\
\bottomrule
\end{tabular}
\end{table}

Three axis-level findings:

\begin{itemize}[leftmargin=*]
  \item \textbf{Specificity saturates.} With code access (code-only or documented), all models achieve specificity $\geq$4.78, indicating that agents reliably reference concrete file paths and function names once they can explore the repository. Specificity is primarily a closed-book discriminator and contributes negligible signal for distinguishing model performance in agentic conditions. Future rubric iterations could replace or supplement specificity with a more discriminative axis such as \emph{integration depth} (how many cross-file connections the answer traces).
  \item \textbf{Completeness benefits most from documentation.} The largest documented$-$code-only gains appear on the completeness axis (+0.53 to +0.75 across models), because documentation provides the missing details (caveats, edge cases, design choices) needed to fully address the rubric.
  \item \textbf{Accuracy improves steadily.} Accuracy rises monotonically from closed-book through code-only to documented, reflecting that more information sources reduce factual errors.
\end{itemize}

\subsection{Per-Category Breakdown}
\label{sec:category-breakdown}

Table~\ref{tab:category} reports scores by question category on code-derivable tasks, testing the hypothesis (Section~\ref{sec:categories}) that \emph{Why} questions show the largest documentation gap.

\begin{table}[h]
\centering
\caption{Per-category scores on code-derivable tasks (pooled across 4 models; 132 tasks per category $\times$ 4 models = 528 observations per cell). $\Delta$ = documented $-$ code-only; $d$ = Cohen's $d$. Bonferroni-adjusted $\alpha$ = 0.0125 (4 tests).}
\label{tab:category}
\begin{tabular}{lcccccc}
\toprule
\textbf{Category} & \textbf{Closed-book} & \textbf{Code-only} & \textbf{Documented} & $\boldsymbol{\Delta}$ & $\boldsymbol{d}$ & $\boldsymbol{p}$ \\
\midrule
What  & 0.484 & 0.860 & 0.863 & $+0.003$ & 0.03 & 0.278 \\
Why   & 0.536 & 0.837 & 0.840 & $+0.003$ & 0.03 & 0.268 \\
Where & 0.479 & 0.863 & 0.874 & $+0.011$ & 0.13 & 0.003 \\
How   & 0.500 & 0.869 & 0.876 & $+0.008$ & 0.06 & 0.069 \\
\bottomrule
\end{tabular}
\end{table}

Contrary to our hypothesis, \emph{Why} questions do \emph{not} show the largest documentation gap on code-derivable tasks ($\Delta$=+0.003, $p$=0.268).
Instead, \emph{Where} questions exhibit the only statistically significant delta ($\Delta$=+0.011, $p$=0.003, $d$=0.13), suggesting that documentation helps most when agents need to locate features across a repository (feature location, dependency tracing), where README files and module docstrings serve as a navigation index.
The negligible gap for \emph{Why} questions is consistent with our design: code-derivable tasks have gold answers grounded entirely in source code, so even design-rationale questions are answerable from code patterns (defensive checks, fallback logic) without requiring documentation.
This validates the task generation pipeline: code-derivable tasks genuinely do not require documentation regardless of question type.

Additionally, \emph{Why} questions have the highest closed-book scores (0.536 vs.\ 0.479--0.500), confirming that design rationale is most susceptible to memorization.
Claude excels on \emph{How} (0.912 code-only) while DeepSeek excels on \emph{What} (0.903), suggesting complementary strengths.

\subsection{Ceiling Effects and Documentation Utility}
\label{sec:ceiling}
\label{sec:doc-utility}

While agents benefit from documentation ($\Delta_{\text{doc}}$ = +0.071 mean on doc-dependent tasks), the gain is modest: agents achieve 0.84--0.87 on code-only, suggesting they infer much documented information from code structure alone.
Documentation provides incremental benefit primarily for completeness---filling in design rationale that cannot be inferred from code patterns.

Three of four models achieve code-only scores above 0.87 on code-derivable tasks, approaching a practical ceiling.
Contributing factors: specificity saturates near 5.0 with repository access, rubric-based scoring rewards key-point coverage that agents with 60 turns reliably achieve, and frontier models are genuinely strong code-structural readers.
The fully automated pipeline supports periodic regeneration from newer repositories, tighter exploration budgets, or deeper architectural questions to maintain discrimination as models improve.

\section{Discussion}
\label{sec:discussion}

\paragraph{What ``code-only'' actually measures.}
\label{sec:code-only-semantics}
The code-only condition preserves identifier names, string literals, imports, and type annotations, all of which carry semantic information (e.g., \texttt{validate\_email\_format} communicates purpose via naming alone).
Thus, we measure \emph{code-structural reasoning} rather than pure syntactic reasoning: the ability to derive understanding from executable source code minus natural-language annotations.
In the program comprehension literature, this maps to a blend of bottom-up comprehension~\citep{pennington1987stimulus} (tracing control flow and data flow from code) and top-down cues from identifier semantics~\citep{schankin2018descriptive}, corresponding to the practical task of reading unfamiliar or undocumented code where developers switch fluidly between strategies~\citep{vonmayrhauser1995program}.
The high code-only scores (0.77--0.89) likely reflect models leveraging descriptive naming conventions alongside control-flow analysis.
This is a deliberate design choice: reading undocumented but well-named code is the practical task we aim to measure.
An obfuscation ablation (replacing identifiers with opaque tokens) would quantify the contribution of naming conventions vs.\ pure structural reasoning; we leave this for future work.
Importantly, the code-access gain (+0.23 mean over closed-book) is almost entirely attributable to accuracy and completeness improvements: a 2-axis composite excluding specificity yields deltas within 3\% of the 3-axis composite (Appendix~\ref{app:per-repo}), confirming that the headline number is not inflated by specificity saturation.

\paragraph{Validation logic and circularity.}
The dual task set creates a falsifiable prediction: code-derivable tasks should show code-only~$\approx$~documented, while doc-dependent tasks should show documented~$>$~code-only.
Both predictions are confirmed (Tables~\ref{tab:code-derivable-results}--\ref{tab:doc-dep-results}).
We acknowledge that code-derivable tasks are \emph{designed} to be code-answerable via the code-only verification pass, so the near-zero delta on these tasks is partially self-confirming.
However, the differential prediction is not circular: the same design predicts a \emph{positive} delta on doc-dependent tasks, which could have failed if documentation were genuinely unhelpful.
The consistent, significant doc-dependent gap across all four models (mean +0.071, $p < 0.003$) provides non-trivial evidence that the benchmark distinguishes the two task types as intended.

\paragraph{Category balance.}
We enforce 25\% per question type via round-robin allocation (Section~\ref{sec:categories}).
While real developer questions are not uniformly distributed---SWE-QA's analysis of 77,100 GitHub questions shows substantial skew---balanced allocation enables equal-powered per-category comparisons and avoids confounding category effects with difficulty.
Results under balanced allocation remain valid for ranking models; users who prefer ecological weighting can re-weight using per-category scores (Table~\ref{tab:category}).

\paragraph{Limitations.}
\label{sec:limitations}
The benchmark has several known limitations.
\emph{Judge reliability:} We use a single LLM judge (GPT-5.4) without inter-judge agreement or human correlation analysis; a reliability study stratified by condition is needed.
\emph{Generator-evaluatee overlap:} Claude generates gold answers and scores highest on some conditions, though the bias is negligible on the primary code-only metric (DeepSeek matches at 0.892 vs.\ 0.891).
\emph{No human validation:} The pipeline relies on automated quality gates; a stratified human evaluation is planned.
The fact-check verification produces 38.4\% pass and 61.6\% warn verdicts; ``warn'' indicates minor discrepancies (e.g., slight imprecision) that do not rise to factual contradiction, and only ``fail'' verdicts are dropped.
While warned tasks are retained, the high warn rate suggests future iterations should tighten the verification threshold or add human review for borderline cases.
\emph{Scope:} The benchmark is Python-only (AST-based documentation removal is language-specific), uses single-commit snapshots, and removal may break tests that inspect \texttt{\_\_doc\_\_} attributes.
\emph{Ceiling effects:} Three of four models score 0.87--0.92 on code-derivable tasks in the code-only condition, with 5--12\% of tasks at ceiling ($\geq$0.95); periodic regeneration or tighter exploration budgets are needed to maintain discrimination as models improve.

\section{Conclusion}
\label{sec:conclusion}

We presented Code-QA-Bench, a fully automated benchmark that separates code comprehension from documentation memorization via a three-condition experimental design and dual task set (528 code-derivable + 100 doc-dependent tasks across 10 repositories).
Experiments on four frontier models show that \textbf{code access is the dominant factor} (+0.23 over closed-book), \textbf{documentation provides consistent but moderate benefit} (+0.071 on doc-dependent tasks, $p < 0.003$), and \textbf{code-only $\approx$ documented on code-derivable tasks} validates the methodology.

The pipeline is repo-agnostic and can be applied to any well-documented Python repository, serving both as a reusable evaluation framework and as a source of verified training data~\citep{cai2025sweqapro}.
The three-condition design complements SWE-QA's~\citep{peng2025sweqa} taxonomic coverage and SWE-QA-Pro's~\citep{cai2025sweqapro} difficulty calibration by providing environment-level control with quantitative contamination analysis.
Code and data are open-source.


\bibliographystyle{plainnat}

\appendix
\section{Example Tasks}
\label{app:examples}

We present four example tasks, one per category, drawn from different repositories.

\paragraph{Where / data\_control\_flow (sympy)}
\begin{quote}
\textbf{Question:} Where does \texttt{ipartfrac} get called, and how does its data flow through \texttt{migcdex}, \texttt{igcdex}, and \texttt{gcdext}?

\textbf{Key files:} \texttt{sympy/functions/elementary/\_trigonometric\_special.py}, \texttt{trigonometric.py}, \texttt{sympy/core/intfunc.py}

\textbf{Rubric (first 3 points):}
\begin{enumerate}
  \item \texttt{ipartfrac} is called in \texttt{cos.\_eval\_rewrite\_as\_sqrt} (not directly in sin's version)
  \item \texttt{ipartfrac} uses \texttt{functools.reduce} with a \texttt{mul} function to compute the product of all denominators
  \item It builds complement values (\texttt{denom // x} for each x) and passes them to \texttt{migcdex}
\end{enumerate}

\textbf{Gold answer (abridged):} The \texttt{ipartfrac} function flows data through the following call chain: it is called in \texttt{cos.\_eval\_rewrite\_as\_sqrt()} in \texttt{trigonometric.py}. Internally it uses \texttt{functools.reduce} to compute total product of denominators, builds complement values, and passes them to \texttt{migcdex} which calls \texttt{igcdex} pairwise...
\end{quote}

\paragraph{How / system\_design (pytest)}
\begin{quote}
\textbf{Question:} How does pytest handle finalizer failures during fixture teardown, and how are multiple exceptions from different teardown stages combined?

\textbf{Key files:} \texttt{src/\_pytest/fixtures.py}, \texttt{src/\_pytest/runner.py}, \texttt{src/\_pytest/skipping.py}

\textbf{Rubric (first 3 points):}
\begin{enumerate}
  \item \texttt{FixtureDef.finish()} runs ALL finalizers even if some fail, collecting exceptions in a list
  \item Finalizers are popped in LIFO order (using \texttt{.pop()})
  \item Multiple exceptions are wrapped in a \texttt{BaseExceptionGroup} with exceptions reversed
\end{enumerate}

\textbf{Gold answer (abridged):} \texttt{FixtureDef.finish()} runs all finalizers even if some fail. Finalizers are popped in LIFO order. If multiple exceptions occur, they are wrapped in a \texttt{BaseExceptionGroup} with \texttt{exceptions[::-1]}. After finalization, \texttt{cached\_result} is set to \texttt{None} and \texttt{\_finalizers.clear()} is called even if finalization fails...
\end{quote}

\paragraph{What / architecture\_exploration (xarray)}
\begin{quote}
\textbf{Question:} What is the role and architecture of \texttt{\_normalize\_path} in xarray's backend system, and where is it used across the codebase?

\textbf{Key files:} \texttt{xarray/backends/common.py}, \texttt{xarray/core/utils.py}, \texttt{xarray/backends/api.py}

\textbf{Rubric (first 3 points):}
\begin{enumerate}
  \item Defined in \texttt{common.py} with three \texttt{@overload} signatures and a single implementation accepting \texttt{os.PathLike | str | T}
  \item Performs two transformations: \texttt{os.fspath()} for PathLike, and \texttt{os.path.abspath(os.path.expanduser(path))} for local strings
  \item Remote URIs detected via \texttt{is\_remote\_uri()} (regex-based in \texttt{utils.py}) are left unmodified
\end{enumerate}

\textbf{Gold answer (abridged):} \texttt{\_normalize\_path} normalizes file paths throughout xarray's backend I/O system. It has three \texttt{@overload} signatures: \texttt{PathLike}$\to$\texttt{str}, \texttt{str}$\to$\texttt{str}, and generic \texttt{T}$\to$\texttt{T}. It converts PathLike objects via \texttt{os.fspath()}, expands local strings, and passes remote URIs through unchanged...
\end{quote}

\paragraph{Why / purpose\_exploration (django)}
\begin{quote}
\textbf{Question:} Why is \texttt{ogrinspect} split into a public function and a private \texttt{\_ogrinspect} generator, and how does the management command exploit that design?

\textbf{Key files:} \texttt{django/contrib/gis/utils/ogrinspect.py}, \texttt{management/commands/ogrinspect.py}, \texttt{gdal/geomtype.py}

\textbf{Rubric (first 3 points):}
\begin{enumerate}
  \item \texttt{\_ogrinspect} is a generator (uses yield) producing model definition lines one at a time; \texttt{ogrinspect} joins them with newlines
  \item The management command calls \texttt{\_ogrinspect} directly to collect lines into a list and append the mapping dictionary before joining
  \item The command uses \texttt{get\_func\_args(\_ogrinspect)} to dynamically filter CLI options to accepted parameters
\end{enumerate}

\textbf{Gold answer (abridged):} The separation serves two purposes: (1) streaming vs.\ string output --- \texttt{\_ogrinspect} yields lines one at a time, allowing the management command to append additional output before joining; (2) dynamic argument filtering --- the command uses \texttt{get\_func\_args} to introspect accepted parameters...
\end{quote}

\section{Documentation Removal Examples}
\label{app:stripping}

The stripping procedure removes all docstrings (via AST), all comments (via tokenize), and all documentation files (README, .md, .rst, docs/ directories). Below are two representative transformations.

\paragraph{Example 1: Docstring and comment removal.}

\begin{minipage}[t]{0.48\textwidth}
\textbf{Before:}
\begin{verbatim}
def calculate_distance(point_a, point_b):
    """Calculate Euclidean distance
    between two points.

    Args:
        point_a: A tuple (x, y).
        point_b: A tuple (x, y).

    Returns:
        The Euclidean distance.
    """
    # Compute squared differences
    dx = point_a[0] - point_b[0]
    dy = point_a[1] - point_b[1]
    return (dx**2 + dy**2) ** 0.5
\end{verbatim}
\end{minipage}
\hfill
\begin{minipage}[t]{0.48\textwidth}
\textbf{After:}
\begin{verbatim}
def calculate_distance(point_a, point_b):
    dx = point_a[0] - point_b[0]
    dy = point_a[1] - point_b[1]
    return (dx**2 + dy**2) ** 0.5
\end{verbatim}
\end{minipage}

\paragraph{Example 2: Docstring-only body replaced with \texttt{pass}.}

\begin{minipage}[t]{0.48\textwidth}
\textbf{Before:}
\begin{verbatim}
class Validator:
    """Base class for all validators."""

    def validate(self, value):
        """Validate the given value.
        Subclasses must override."""
        raise NotImplementedError

    def get_help_text(self):
        """Return a description of what
        this validator checks."""
\end{verbatim}
\end{minipage}
\hfill
\begin{minipage}[t]{0.48\textwidth}
\textbf{After:}
\begin{verbatim}
class Validator:

    def validate(self, value):
        raise NotImplementedError

    def get_help_text(self):
        pass
\end{verbatim}
\end{minipage}

\vspace{0.5em}
Note that \texttt{get\_help\_text} had only a docstring as its body, so the stripper inserts \texttt{pass} to maintain syntactic validity. The \texttt{validate} method retains its \texttt{raise} statement because it was a real code statement, not just documentation. Identifier names, type annotations, string literals, and import statements are preserved unchanged.

\section{Judge Prompt}
\label{app:judge-prompt}

The full prompt given to the LLM judge:

\begin{verbatim}
You are an expert judge evaluating an AI agent's answer about a code repository.

## Question
{question}

## Reference Answer (Gold)
{gold_answer}

## Key Points the Answer Should Cover (Rubric)
{rubric}

## Agent's Answer
{agent_answer}

## Instructions
Score the agent's answer on three axes, each from 0 to 5:

1. Accuracy (0-5): Are the factual claims correct?
2. Completeness (0-5): Does the answer cover all rubric points?
3. Specificity (0-5): Does the answer reference specific files/functions?

Return JSON: {"accuracy": N, "completeness": N, "specificity": N,
              "explanation": "..."}
\end{verbatim}

\section{Comparison with SWE-QA and SWE-QA-Pro}
\label{app:comparison}

\begin{table}[h]
\centering
\caption{Feature comparison of repository-level code QA benchmarks.}
\label{tab:comparison}
\small
\begin{tabular}{p{0.22\textwidth}p{0.22\textwidth}p{0.22\textwidth}p{0.24\textwidth}}
\toprule
\textbf{Feature} & \textbf{SWE-QA} & \textbf{SWE-QA-Pro} & \textbf{Code-QA-Bench} \\
\midrule
Repositories & 12 (popular) & 26 (long-tail) & 10 (popular, 3 cond.) \\
Questions & 576 & 260 & 528 + 100 doc-dep \\
Taxonomy & 4-type, 12 sub & 4-type, 12 sub & 4-type, 12 sub (balanced) \\
Question source & GitHub issues & Issue clusters & Doc chunks \\
Doc in eval & Full & Full & Code-only / Documented / None \\
Anti-memorization & None & Difficulty calibration & 3-condition design \\
Answer generation & RAG + human & Claude Code + human & Answer-first agent \\
Human validation & Yes & Yes & Automated (LLM) \\
Scoring axes & 5 & 5 & 3 \\
Eval conditions & 1 & 1 & 3 \\
Training data & No & Yes (SFT+RL) & Pipeline supports \\
\bottomrule
\end{tabular}
\end{table}

\section{Code-Only Verification Prompt}
\label{app:strip-verify-prompt}

The code-only verification pass (Section~\ref{sec:strip-verify}) uses a single LLM call with the following prompt structure.
The system prompt establishes the auditor role:

\begin{verbatim}
You are an auditor for a code-understanding benchmark. Your job is to
identify "doc-leakage" -- claims in a gold answer that come from
documentation but are NOT recoverable by reading the code-only source
code alone.

Stripped code has ALL docstrings, comments, and documentation files
removed. Only bare Python source remains: function/class definitions,
logic, imports.
\end{verbatim}

The user prompt presents the gold answer and the code-only versions of the key files:

\begin{verbatim}
Gold answer:
--- answer ---
{answer text}
--- end answer ---

Key files referenced (stripped versions shown below):
--- path/to/file.py ---
{stripped source code}
--- end path/to/file.py ---

For each factual claim in the gold answer, decide:
- KEEP: verifiable from code-only code (function names, class
  hierarchies, imports, control flow, algorithm logic)
- REMOVE: relies on documentation (design rationale from docstrings,
  purpose descriptions from README, parameter semantics)
- REWRITE: partially verifiable -- keep the code-grounded part

Respond with JSON:
{
  "claims": [{"claim": "...", "verdict": "keep|remove|rewrite",
              "reason": "..."}],
  "revised_answer": "answer with REMOVE claims deleted",
  "doc_leakage_fraction": 0.0 to 1.0,
  "summary": "one-line summary"
}
\end{verbatim}

The key files are stripped using the same AST-based procedure as the evaluation repository (Section~\ref{sec:doc-removal}), ensuring the auditor sees exactly the code that agents will encounter during evaluation.

\section{Answer Generation Prompt}
\label{app:answer-prompt}

The answer generation agent receives two prompts. The \textbf{system prompt} establishes the agent's role and exploration strategy:

\begin{verbatim}
You are generating a gold-standard answer for a code-understanding
benchmark. You have tools to explore the repository: read_file,
list_directory, search_code.

Your answer will be used as the ground truth to judge AI agents. It
must contain specific, verifiable facts that can ONLY be known by
reading the actual source code.

Exploration strategy:
1. Start by locating the code referenced in the documentation chunk
2. Read the primary file(s) and identify the key functions/classes
3. Go ONE LEVEL DEEPER: trace at least one callee, one parent class,
   or one related module to understand how the code connects to the
   broader system
4. Your answer must include facts you discovered from code that are
   NOT stated in the documentation chunk -- this is what makes the
   benchmark challenging
\end{verbatim}

The \textbf{user prompt} provides the documentation chunk as a topic guide and specifies the target category:

\begin{verbatim}
The following documentation chunk from "{repo_name}" identifies the
TOPIC for your answer. Use it to know WHAT to investigate -- but your
answer must go beyond it by reading actual code.

--- chunk ---
[{source}: {file_path}] {heading}
{content}
--- end chunk ---

Target question type: "{category}" (sub-type: "{sub_type}")

Exploration guidance by category:
- "what": Read the module structure, class hierarchies, and imports
  to map architecture and dependencies
- "why": Look for code patterns that reveal design decisions --
  defensive checks, performance optimizations, fallback logic
- "where": Trace the call chain: who calls this, what does it call,
  where does data flow through
- "how": Read the implementation line by line -- understand the
  algorithm, the state transitions, the edge case handling

Your job:
1. Use tools to find and read the source code
2. Go deeper: trace at least one callee, parent class, or import
3. Write an answer that includes specific code details NOT found in
   the documentation chunk above

When done exploring, respond with JSON (no tool calls):
{
  "answer": "...",
  "key_files": ["relative/path.py", ...],
  "code_evidence": [
    "specific fact verified by reading code (file + function)", ...
  ]
}

Your answer MUST include at least 3 items in code_evidence.
\end{verbatim}

\section{Question Generation Prompt}
\label{app:question-prompt}

The question generation agent derives a natural question from the gold answer. The \textbf{system prompt}:

\begin{verbatim}
You are writing questions for a code-understanding benchmark. The
questions should sound like a curious developer asking a colleague --
short, natural, and conversational.

Rules:
- One sentence, max ~25 words
- Do NOT mention file paths, directory names, or line numbers
- Do NOT use phrases like "provide a comprehensive analysis",
  "describe in detail", "trace through", "for each step identify"
- Use natural language: "What happens when...", "How does X work?"
- The agent being tested will see code with ALL docstrings, comments,
  and docs removed
- You have tools to read files, list directories, and search code
\end{verbatim}

The \textbf{user prompt} provides the gold answer and category guidance:

\begin{verbatim}
Gold answer (what the correct response should cover):
{answer}

Target question type: "{category}" (sub-type: "{sub_type}")

Write a short, natural "{category}" question that this answer would
correctly respond to.

Category guidance -- your question MUST match the target type:
- "what": Ask about structure, definition, or relationships.
  Examples: "What components make up the caching subsystem?"
- "why": Ask about rationale, purpose, or design decisions.
  Examples: "Why does the ORM use lazy evaluation?"
- "where": Ask about location, data flow, or identifiers.
  Examples: "Where is the request routing logic implemented?"
- "how": Ask about implementation, algorithms, or operation.
  Examples: "How does the template engine resolve lookups?"

Bad examples (DO NOT write like these):
- "Trace through function X in file Y and describe in detail..."
- "Provide a comprehensive architectural analysis of..."

The question must be answerable ONLY by reading source code (no docs).

Respond with JSON:
{
  "question": "...",
  "rubric": ["point1", ...]
}
\end{verbatim}

\section{Bootstrap Confidence Intervals}
\label{app:bootstrap}

All means and deltas reported in the main text are accompanied by 95\% bootstrap confidence intervals computed via 10{,}000 resamples with replacement (seed = 42).
For paired comparisons (e.g., documented$-$code-only), we use paired bootstrap: resampling task indices jointly and computing the delta on each resample.
$p$-values are computed as the proportion of bootstrap samples where the delta crosses zero.

\begin{table}[h]
\centering
\caption{95\% bootstrap CIs for code-derivable task scores (528 tasks). CI width of $\pm$0.006--0.014 reflects tight estimation.}
\label{tab:bootstrap-cd}
\begin{tabular}{lccc}
\toprule
\textbf{Model} & \textbf{Closed-book} & \textbf{Code-only} & \textbf{Documented} \\
\midrule
Claude Opus 4.6 & 0.560 \scriptsize{[.548, .571]} & 0.891 \scriptsize{[.885, .897]} & 0.918 \scriptsize{[.912, .922]} \\
DeepSeek-V4-Pro & 0.442 \scriptsize{[.430, .455]} & 0.892 \scriptsize{[.884, .899]} & 0.899 \scriptsize{[.893, .905]} \\
Kimi-K2.6 & 0.514 \scriptsize{[.501, .527]} & 0.873 \scriptsize{[.867, .880]} & 0.882 \scriptsize{[.874, .889]} \\
Gemini-3.1-Pro & 0.482 \scriptsize{[.468, .497]} & 0.772 \scriptsize{[.761, .784]} & 0.755 \scriptsize{[.742, .766]} \\
\bottomrule
\end{tabular}
\end{table}

\begin{table}[h]
\centering
\caption{95\% bootstrap CIs for doc-dependent task scores (100 tasks). Wider CIs reflect smaller sample size.}
\label{tab:bootstrap-dd}
\begin{tabular}{lccc}
\toprule
\textbf{Model} & \textbf{Closed-book} & \textbf{Code-only} & \textbf{Documented} \\
\midrule
Claude Opus 4.6 & 0.682 \scriptsize{[.643, .719]} & 0.873 \scriptsize{[.847, .898]} & 0.953 \scriptsize{[.941, .965]} \\
DeepSeek-V4-Pro & 0.557 \scriptsize{[.514, .597]} & 0.867 \scriptsize{[.833, .898]} & 0.928 \scriptsize{[.903, .947]} \\
Kimi-K2.6 & 0.639 \scriptsize{[.597, .681]} & 0.859 \scriptsize{[.825, .890]} & 0.950 \scriptsize{[.937, .962]} \\
Gemini-3.1-Pro & 0.636 \scriptsize{[.592, .679]} & 0.840 \scriptsize{[.808, .871]} & 0.893 \scriptsize{[.864, .917]} \\
\bottomrule
\end{tabular}
\end{table}

\begin{table}[h]
\centering
\caption{Paired bootstrap test for $\Delta_{\text{doc}}$ (documented $-$ code-only). Code-derivable deltas are near zero for most models; doc-dependent deltas are consistently positive and significant.}
\label{tab:bootstrap-deltas}
\begin{tabular}{llccc}
\toprule
\textbf{Task Set} & \textbf{Model} & $\boldsymbol{\Delta}$ & \textbf{95\% CI} & $\boldsymbol{p}$ \\
\midrule
\multirow{4}{*}{Code-deriv.}
  & Claude Opus 4.6 & +0.026 & [+0.021, +0.033] & $<$0.001 \\
  & DeepSeek-V4-Pro & +0.007 & [$-$0.001, +0.016] & 0.039 \\
  & Kimi-K2.6 & +0.008 & [$-$0.001, +0.017] & 0.033 \\
  & Gemini-3.1-Pro & $-$0.018 & [$-$0.031, $-$0.005] & 0.004 \\
\midrule
\multirow{4}{*}{Doc-dep.}
  & Claude Opus 4.6 & +0.080 & [+0.059, +0.103] & $<$0.001 \\
  & Kimi-K2.6 & +0.091 & [+0.062, +0.123] & $<$0.001 \\
  & DeepSeek-V4-Pro & +0.061 & [+0.024, +0.099] & $<$0.001 \\
  & Gemini-3.1-Pro & +0.053 & [+0.015, +0.091] & 0.002 \\
\bottomrule
\end{tabular}
\end{table}

The code-derivable results validate the code-only verification design: for DeepSeek and Kimi, the documented$-$code-only CI includes zero, confirming that gold answers are code-recoverable.
Claude's positive delta (+0.026, $p < 0.001$) likely reflects generator advantage rather than documentation dependency (Section~\ref{sec:experiments}).
Gemini's negative delta ($-$0.018, $p = 0.004$, marginally significant) suggests that documentation occasionally distracts from code-grounded reasoning for this model.

For doc-dependent tasks, all models show significant positive deltas (minimum $p = 0.002$), confirming that documentation utility is real and measurable.
The wider CIs (reflecting $N=100$) mean that model-to-model differences in $\Delta_{\text{doc}}$ are not individually significant, but the consistent directionality across all four models provides strong evidence for the aggregate effect.

\section{Per-Repository Breakdown}
\label{app:per-repo}

Table~\ref{tab:per-repo} shows code-only scores by repository for all four models.
Scores are consistent across repositories: within each model, the range spans only 0.04--0.07 points, confirming that aggregate results are not driven by a subset of easy or hard repositories.

\begin{table}[h]
\centering
\caption{Per-repository scores in the code-only condition on code-derivable tasks. Range = max $-$ min within each model.}
\label{tab:per-repo}
\small
\begin{tabular}{lcccc}
\toprule
\textbf{Repository} & \textbf{Claude} & \textbf{DeepSeek} & \textbf{Kimi} & \textbf{Gemini} \\
\midrule
django       & 0.889 & 0.881 & 0.876 & 0.786 \\
pylint       & 0.889 & 0.894 & 0.876 & 0.783 \\
sympy        & 0.886 & 0.892 & 0.868 & 0.787 \\
scikit-learn & 0.885 & 0.886 & 0.879 & 0.750 \\
astropy      & 0.893 & 0.906 & 0.879 & 0.736 \\
matplotlib   & 0.893 & 0.904 & 0.881 & 0.793 \\
sphinx       & 0.897 & 0.890 & 0.844 & 0.751 \\
pytest       & 0.889 & 0.894 & 0.868 & 0.739 \\
xarray       & 0.907 & 0.904 & 0.886 & 0.792 \\
seaborn      & 0.886 & 0.850 & 0.883 & 0.803 \\
\midrule
Range        & 0.022 & 0.056 & 0.042 & 0.067 \\
\bottomrule
\end{tabular}
\end{table}

\end{document}